\documentclass[prl,aps,a4paper,oneside,superscriptaddress,preprint,showpacs,showkeys]{revtex4}

\usepackage{graphicx}  
\usepackage{latexsym}   

\begin{document}

\title{Optimum exploration memory and anomalous diffusion in deterministic partially self-avoiding walks in one-dimensional random media}

\author{\firstname{C\'esar} Augusto Sangaletti \surname{Ter\c{c}ariol}} 
\email{cesartercariol@gmail.com}
\affiliation{Faculdade de Filosofia, Ci\^encias e Letras de Ribeir\~ao Preto, \\
             Universidade de S\~ao Paulo \\ 
             Avenida Bandeirantes, 3900 \\ 
             14040-901, Ribeir\~ao Preto, SP, Brazil.}

\affiliation{Centro Universit\'ario Bar\~ao de Mau\'a \\
             Rua Ramos de Azevedo, 423 \\ 
             14090-180, Ribeir\~ao Preto, SP, Brazil }

\author{\firstname{Rodrigo} Silva \surname{Gonz\'alez}}
\email{caminhos_rsg@yahoo.com.br}
\affiliation{Faculdade de Filosofia, Ci\^encias e Letras de Ribeir\~ao Preto, \\
             Universidade de S\~ao Paulo \\ 
             Avenida Bandeirantes, 3900 \\ 
             14040-901, Ribeir\~ao Preto, SP, Brazil.}

\author{\firstname{Alexandre} Souto \surname{Martinez}}
\email{asmartinez@ffclrp.usp.br}

\affiliation{Faculdade de Filosofia, Ci\^encias e Letras de Ribeir\~ao Preto, \\
             Universidade de S\~ao Paulo \\ 
             Avenida Bandeirantes, 3900 \\ 
             14040-901, Ribeir\~ao Preto, SP, Brazil.}
\date{\today}

\begin{abstract}
Consider $N$ points randomly distributed along a line segment of unitary length.
A walker explores this disordered medium moving according to a partially self-avoiding deterministic walk.
The walker, with memory $\mu$, leaves from the leftmost point and moves, at each discrete time step, to the nearest point, which has not been visited in the preceding $\mu$ steps.
We have obtained analytically the probability $P_N(\mu) = (1 - 2^{-\mu})^{N - \mu - 1}$ that all $N$ points are visited in this open system, with  $N \gg \mu \gg 1$.
The expression for $P_N(\mu)$ evaluated in the mentioned limit is valid even for small $N$ and leads to a transition region 
centered at $\mu_1 = \ln N/\ln 2$ and with width $\varepsilon = e/\ln2$. 
For $\mu < \mu_1 - \varepsilon/2$, the walker gets trapped in cycles and does not fully explore the system. 
For $\mu > \mu_1 + \varepsilon/2$ the walker explores the whole system.  
In both cases the walker presents diffusive behavior.
Nevertheless, in the intermediate regime $\mu \sim \mu_1 \pm \epsilon/2$, the walker presents anoumalous diffusion behavior. 
Since the intermediate region increases as $\ln N$ and its width is constant, a sharp transition is obtained for one-dimensional large systems.
The walker does not need to have full memory of its trajectory to explore the whole system, 
it suffices to have memory of order $\mu_1$.  
\end{abstract}



\maketitle


While random walks in regular or disordered media have been thouroughly explored~\cite{fisher:1984}, deterministic walks in regular~\cite{grassberger:92} and disordered media~\cite{bunimovich:2004,boyer_2004,boyer_2005,boyer_2006} have been much less studied. 
Here we are concerned with the properties of deterministic walks in random media.

Given $N$ points (cities) distributed in a $d$-dimensional space, a possible question to ask is how efficiently these cities can be visited by a walker who follows a simple movimentation rule. 
In the \emph{travelling salesman problem}, one searches for the shortest closed path, which passes once in each city. 
This problem has been extensively studied. 
In particular, if the coordinates of the cities are distributed following a uniform deviate, results concerning the statistics of the shortest paths have been obtained analytically~\cite{percus:1996, percus:1997, percus:1999}. 
To tackle this problem, one imperatively needs to know the coordinates of all the cities in advance.
Global system information must be at the walker's disposal.

Nevertheless, other situations may be envisaged.
For instance, suppose that only local information about the neighborhood ranking of the current city is at the walker's disposal.
In this case, one can think of several deterministic and stochastic strategies to maximize the number of visited cities, while trying to minimize the travelled distance. 

Our aim is to study the way a walker explores the medium following the deterministic rule of going to the nearest point, which has not been visited in the previous $\mu$ discrete time steps. 
We call this partially self-avoiding walk of the \emph{deterministic tourist walk}~\cite{lima_prl2001,stanley_2001,kinouchi:1:2002}.
Each trajectory, produced by this deterministic rule, has an initial transient of length $t$ and ends in a cycle of period $p$. 
Both transient time and cycle period can be combined in the joint distribution $S_{\mu, d}^{(N)}(t,p)$. 
The $\mu = 0$ deterministic tourist walk is trivial, the walker does not move at each time step. 
The transient and period joint distribution is simply $S^{(N)}_{0, d}(t, p) = \delta_{t,0} \delta_{p,1}$, where $\delta_{i,j}$ is the Kronecker delta. 
With memory $\mu = 1$, the walker must leave the current city at each time step and the transient time and period joint distribution has been obtained for $N \gg 1$~\cite{tercariol_2005}. 
The $\mu = 1$ walks do not lead to the random medium exploration. 
This is due to very short transient times and the tourist gets trapped in pairs of cities, which are mutually nearest neighbors. 
Interesting phenomena occur when $\mu \ge 1$ is considered. 
In this case, the cycle distribution is no longer peaked at $p_{min} = \mu + 1$, but presents a whole spectrum of cycles with period $p \in [\mu+1, N]$, with possible power-law decay~\cite{lima_prl2001,stanley_2001,kinouchi:1:2002}. 
These cycles have been used as a clusterization method~\cite{campiteli_2006}, texture analysis in images~\cite{backes_2006,bruno_2006}, primate foraging~\cite{boyer_2006,boyer_2004} and thesaurus analysis~\cite{kinouchi:1:2002}. 

It is interesting to point out that, for 1D systems, determinism imposes serious restrictions.
For any $\mu$ value, cycles with period $p \in [2\mu+1, 2\mu+3]$ are forbidden.
Additionally, for $\mu = 2$ all odd periods, but $p_{min} = 3$, are forbidden.
Also, the heavy tail of the period marginal distribuition $S_{\mu, 1}^{(N)}(p) = \sum_t S_{\mu, 1}^{(N)}(t,p)$ may lead to often-visited-large-period cycles~\cite{lima_prl2001}.
This allows system exploration even for small memory values ($\mu \ll N$).

In this letter, we consider the deterministic tourist walk along 1D random systems. 
Through a simple (though not trivial) derivation, we show the existence of a transition in the walker's exploratory behavior at a critical memory $\mu_1 = \ln N / \ln 2$ in a narrow memory range of width $\varepsilon = e/\ln 2$. 
This transition splits the walker's behavior in essentially three regimes.
Also, we show numerically that the final step distribution of these walks is the estimator of the fractionary and non-linear diffusion equation solution~\cite{metzler_2000,anteneodo_2005}. 
For $\mu < \mu_1 - \varepsilon/2$, the walker gets trapped in cycles and for $\mu > \mu_1 + \varepsilon/2$, the walker visits all the cities.
In both cases, the walker presents normal diffusion behavior. 
Nevertheless, for $\mu \sim \mu_1 \pm \varepsilon/2$ the walker superdiffuses indicating that $\mu_1$ is an optimum memory for maximum exploration and minimum movimentation.


A random static semi-infinite medium is constructed by uncountable points that are randomly and uniformly distributed along a line segment with a mean point density $r$.
The distances $x$ between consecutive points follow an exponential probability density function (pdf): $f(x) = r e^{-r x}$, for $x \ge 0$ and $f(x) = 0$, otherwise.

Consider now the tourist dynamics with a walker who leaves from the city $s_1$, placed at the origin of the line segment.
The probability $S_{\mu, si}^{(\infty)}(n)$ for the walker to explore $n$ distinct cities can be derived as follows.
The first $\mu+1$ cities are indeed explored, because the memory $\mu$ prohibits the walker to turn back.
Thus, the distances $x_1$, $x_2$, \ldots, $x_\mu$ may assume any value in the interval $[0, \infty)$.
The following steps are uncertain.
The walker may move either forward to a new city or backward to an already visited city outside the memory window.
In analogy to the geometric distribution, it is useful to define the exploration probability $\tilde{q}_j$ as the probability for the walker to explore a new city at the $j$-th uncertain step. 
Thus, $\tilde{q}_1$ can be obtained imposing that the distance $x_{\mu+1}$ must be less than the sum $y_1=\sum_{k=1}^\mu x_k$.
Since the variables $x_1$, $x_2$, \ldots, $x_\mu$ are independent and identically distributed with exponencial pdf, $y_1$ has a gamma pdf.
Hence $\tilde{q}_1 = \int_0^\infty dy_1 r^\mu y_1^{\mu-1} e^{-ry_1} /\Gamma(\mu) \int_0^{y_1} dx_{\mu+1} re^{-rx_{\mu+1}} = 1-2^{-\mu}$. 

The exploration probability $\tilde{q}_2$ for the second uncertain step is not exactly equal to $\tilde{q}_1$.
Once the distance $x_{\mu+1}$ must vary in the interval $[0, y_1]$, the variables $x_2$, $x_3$, \ldots, $x_{\mu+1}$ are not all independent, and consequently $y_2=\sum_{k=2}^{\mu+1} x_k$ has not exactly a gamma pdf.
However, for $\mu \gg 1$, $x_{\mu+1}$ rarely exceeds $y_1$ [this probability is just $P(x_{\mu+1}>y_1) = 1-\tilde{q}_1=2^{-\mu}$, meaning that a weak correlation is present].
Therefore, one can make an approximation assuming that $y_2$ follows a gamma pdf and considering $\tilde{q}_2 \approx \tilde{q}_1$.
The same argumentation can be used for the succeeding steps.

When the city $s_n$ is reached, the walker must turn back, finishing the medium exploration.
Once $\tilde{q_1}$ is taken for all $\tilde{q}$, the return probability is $\tilde{p} = 1-\tilde{q} = 2^{-\mu}$ and one has: $S_{\mu, si}^{(\infty)}(n) = 2^{-\mu} (1-2^{-\mu})^{n-\mu-1}$.
Notice that $r$ has been eliminated, indicating that the number of explored cities does not depend on the medium density.

We observe that once the memory $\mu$ assures the walker to visit at least $\mu+1$ cities, its convenient to define the number of extra explored cities as $n_e = n-\mu-1$. 
In this variable, $S_{\mu, si}^{(\infty)}$ begins at $n_e=0$ for all $\mu$ value and one has $\mbox{E}(n_e) = 2^\mu-1$, which may be interpreted as the characteristic reach of the walk, and $\mbox{Var}(n_e) = 2^{2\mu}-2^\mu$. 

The finite disordered medium is constructed by $N$ points, whose coordinates are randomly (uniform deviate) generated in the interval $[0, 1]$.
The exploration and return probabilities obtained for the semi-infinite medium may also be applied to this finite medium.
The equivalence between these two media can be shown restricting the semi-infinite medium length to the first $N$ points and normalizing it to fit in the interval $[0, 1]$.
For both media, the abscissas of the ranked points follow a beta pdf.

\begin{figure}[htb]
\begin{center}
\includegraphics[angle=-90,width=.7\columnwidth]{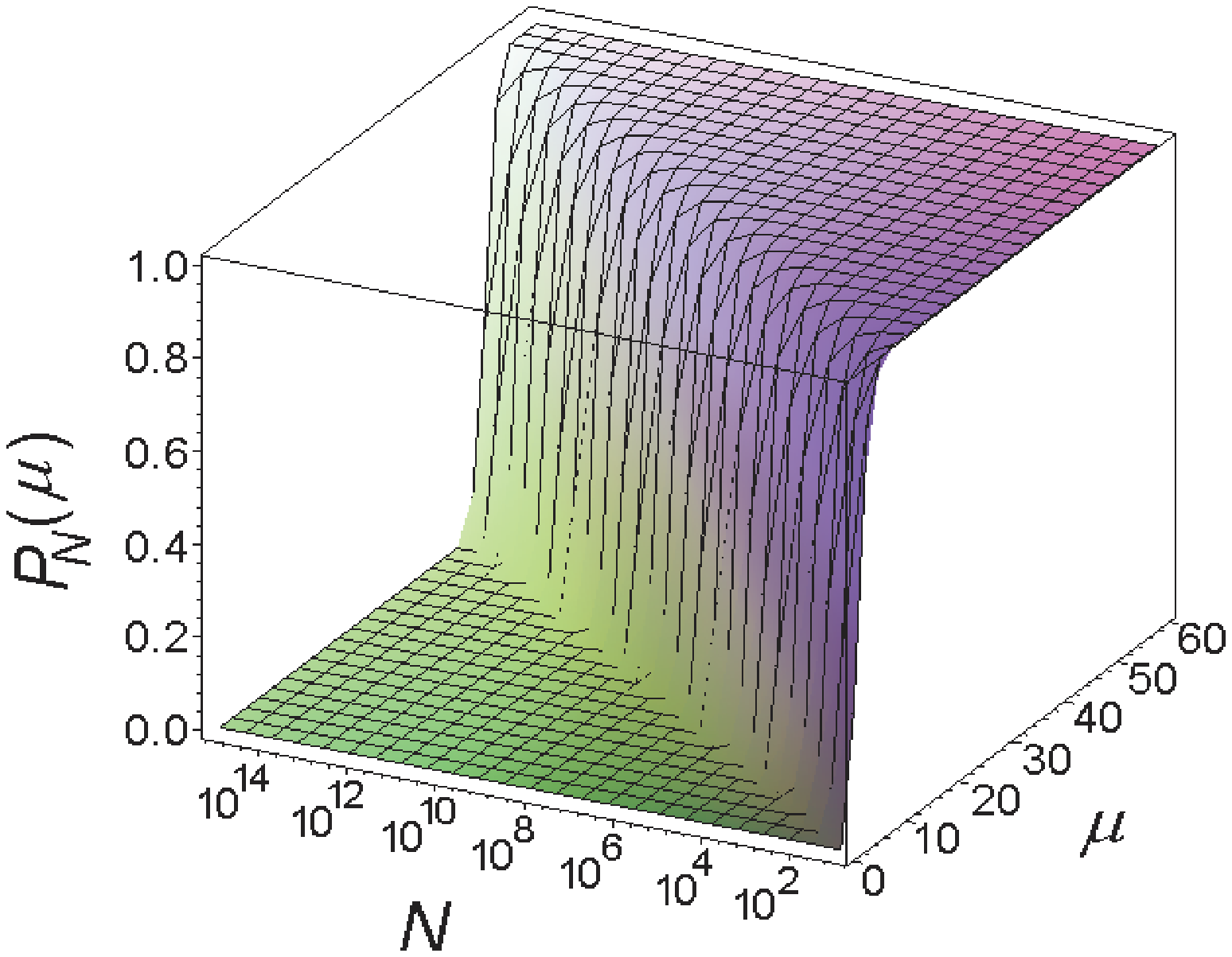}
{\bf (a)}
\includegraphics[angle=-90,width=.7\columnwidth]{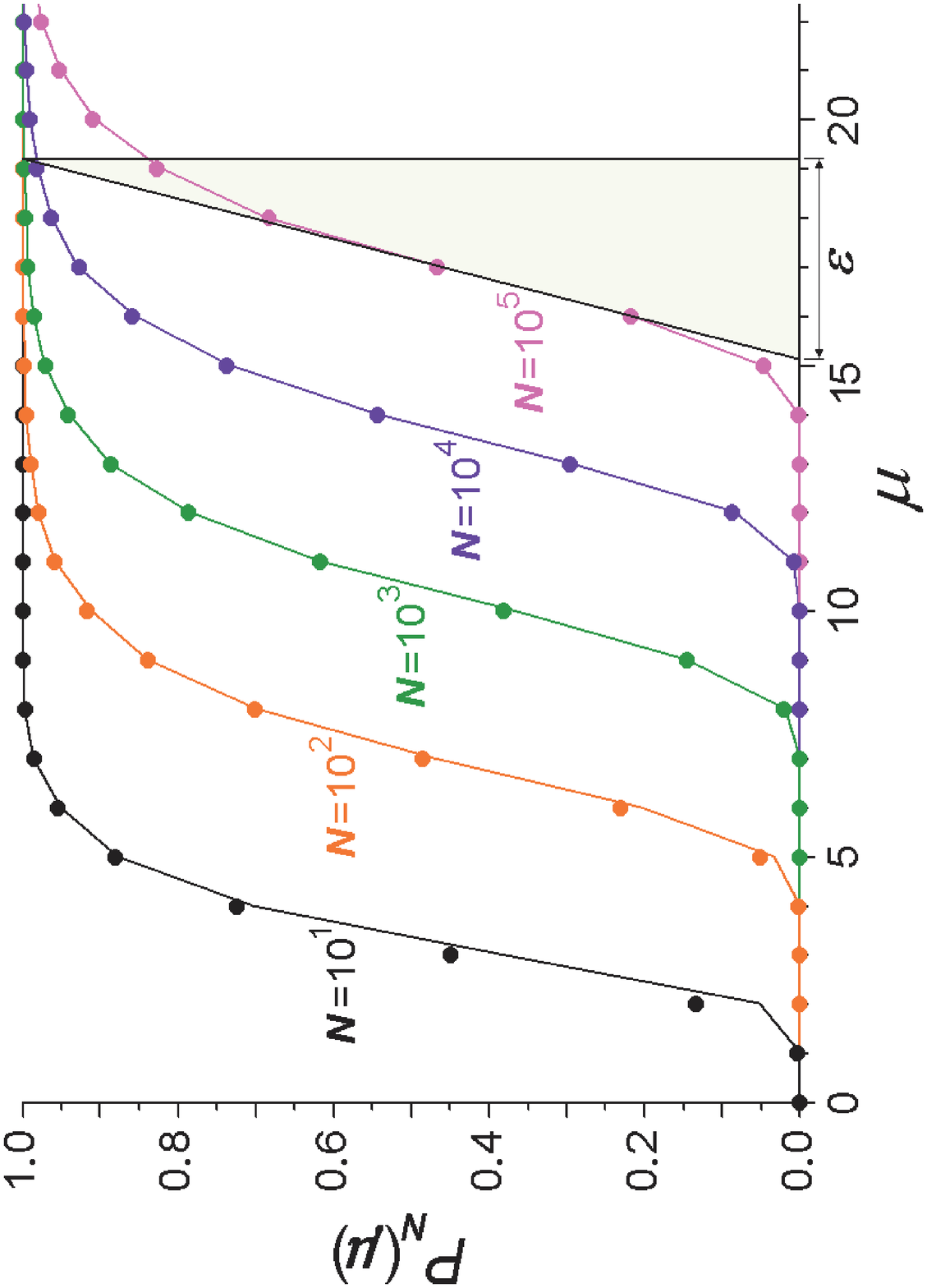}
{\bf (b)}
\caption{
{\bf (a)} Probability of full exploration (Eq.~\ref{Eq:Percolacao}) as a function of $\mu$ and $N$ for 1D deterministic tourist walks.
One sees the abrupt transiton between the localized (bottom) and extended (upper) regimes. 
{\bf (b)} Projections of Eq.~\ref{Eq:Percolacao} for some fixed $N$ values. 
Analytical results are satisfactory, when compared to numerical simulation ($M=10\,000$ maps for each $N$ and $\mu$ values), even for small $N$ and $\mu$ values. 
Error bars are smaller than symbol size.
The transition points $\mu_1$ are given by Eq.~\ref{Eq:MemCritica}, which are weakly dependent on $N$ but all of them have the same constant dispersion $\varepsilon \sim 4$ (Eq.~\ref{Eq:RegTrans}).}
\label{Fig:Percolacao}
\end{center}
\end{figure}

The probability $P_N(\mu)$ for the exploration of the whole $N$-point medium can be derived noticing that the walker must move forward $N-(\mu+1)$ uncertain steps and, when the last city $s_N$ is reached, there is no need to impose a return step.
Therefore
\begin{eqnarray}
\label{Eq:Percolacao}
P_N(\mu) = \tilde{q}^{\, n_e} = \left(1-2^{-\mu}\right)^{N-\mu-1} \; ,
\end{eqnarray}
which is plotted in Fig.~\ref{Fig:Percolacao}(a).

In Fig.~\ref{Fig:Percolacao}(b) one sees that the probability of full medium exploration increases rapidly from 0 to 1, in a well defined transition region.
Considering $N \gg \mu$, Eq.~\ref{Eq:Percolacao} may be approximated to $P_N(\mu) = (1-2^{-\mu})^N$.
To obtain the critical memory $\mu_1$, one considers the inflexion point, leading to
\begin{eqnarray}
\label{Eq:MemCritica}
\mu_1 & = & \log_2 N \; ,
\end{eqnarray}
which is the number of bits to represent the system size. 
For all $N$, the curve slope at $\mu_1$ is $\ln 2/e$  [Fig.~\ref{Fig:Percolacao}(b)], the transition region has a constant width
\begin{eqnarray}
\label{Eq:RegTrans}
\varepsilon = \frac{e}{\ln 2} \approx 3.92 \; .
\end{eqnarray}
This indicates that as $N$ increases, $\mu_1$ slowly increases, but its deviation is independent of $N$, so that a sharp transition is found for $N \gg 1$.
The comparison between the analytical result and computer simulation is depicted in Fig.~\ref{Fig:Percolacao}(b). 
Observe that the approximation employed leads to satisfactory results even for small $N$.  

Up to now, we have studied the deterministic walks in 1D disordered media with open boundary condition.
Now, let us focus on the diffusion process.
This tells us about the medium exploration bulk characteristics. 
Using periodic boundary conditions for each map, the walkers start to move, following the tourist rule, from the most central point of the interval $[-1/2, 1/2]$.
After $N$ steps, the walkers enter the periodic part, which can be interpreted as the steady state of the exploratory behavior (no new visited cities).

\begin{figure}[htb]
\begin{center}
\includegraphics[width=.6\columnwidth, angle=-90]{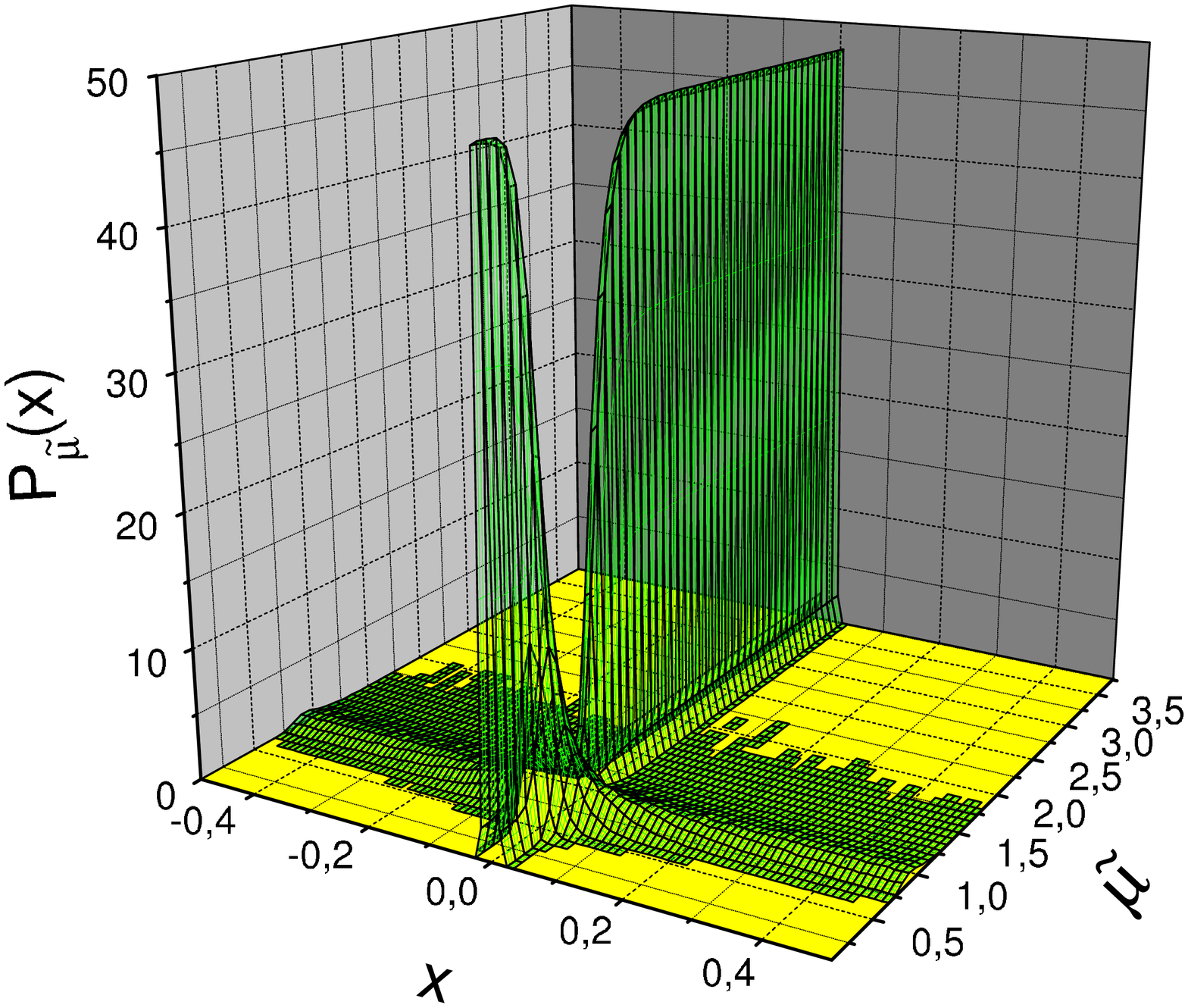}
\caption{Distribution of the final position after $t = N$ steps in the 1D tourist walk as a function of $\mu$.
Numerical simulations have been perfomed with $\Delta x = 1/100$, $M = 10^5$ maps with $N = 10^4$ cities each using periodic boundary conditions. 
The memory $\mu$ varied from $1$ to $50$.}
\label{Fig:SupNormalizada}
\end{center}
\end{figure}

For a given reduced memory $\tilde{\mu} = \mu/ \mu_1$ and after $t=N$ steps, the final position pdf $\rho(x, t)$ has been estimated by $P_{\tilde{\mu}}(x) = n(x)/(M \Delta x)$, where $M$ is the number of maps and $n(x)$ is the number of travelers with final position between $x$ and $x+\Delta x$.
The normalized distribution is shown in Fig.~\ref{Fig:SupNormalizada}, where one sees the huge dispersion around  $\tilde{\mu} = 1$. 

We have shown numerically that $P_{\tilde{\mu}}(x) \sim \rho(x, t)$ is the solution of the fractionary and non-linear difusion equation~\cite{metzler_2000,anteneodo_2005}:
\begin{equation}
\label{Eq:Difusao}
\partial_t \rho(x, t) = D \partial_x^{2+\xi} \rho^{2-q}(x, t) \; .
\end{equation}
Here $\xi$ is the fractionality parameter, which is related to the step-length pdf second-moment divergence, $q$ is the non-linearity parameter, which expresses the influence of correlation between step lengths and $D$ is the diffusion coeficient. 

For $|\mu-\mu_1| > \varepsilon/2$, $\rho(x, t)$ has gaussian ($\xi=0$, $q=1$) shapes indicating normal diffusion.
Further, as $|\mu-\mu_1|$ increases, the variance $\sigma^2$ tends to zero and $\rho(x, t)$ becomes a Dirac delta function: $S_{\mu = 1, 1}^{(N)}(p) = \delta_{p,2}$ or $S_{\mu > \mu_1, 1}^{(N)}(p) \approx \delta_{p,N}$.

Anomalous diffusion occurs for $|\mu - \mu_1| < \varepsilon/2$.
For $|\mu - \mu_1|$ smaller, but closer to $\varepsilon/2$, the walks have finite second moment. 
The $q$ parameter becomes important and the solutions of Eq.~\ref{Eq:Difusao}, with $\xi=0$, are the generalized $q$-gaussians~\cite{anteneodo_2005}: $G_{q}(x) = \gamma_q e_q(-x^2/a_q^2)/\sqrt{a_q^2 \pi}$, where $e_q(x) = [1+(1-q)x]^{1/(1-q)}$ is a generalization for the exponential function, $a_q^2(t) = [2Dt (2 - q)(3 - q) \gamma^{q-1}]^{2/(3 - q)}$ is related to the variance [$\sigma_q^2 = a_q^2/(5-3q)$] of $\rho$ and normalization is possible only for $q<3$:
$\gamma_q = \sqrt{|q-1|} \{ \Gamma[ 1/(q-1) - \delta] / \Gamma \{ (3-q)/[2(q-1)] - \delta \} \}^{1-2\delta}$,   
where $\delta=1$ for $q<1$, which expresses subdiffusion and $\delta=0$ for $1 \le q < 3$ expressing superdiffusion. 
If $q \rightarrow 1^+$, then $\gamma \rightarrow 1$, and one has normal diffusion. 
The variance is finite only for $q < 5/3$, otherwise the $q$-gaussians give place to L\'{e}vy $\alpha$-stable pdf ($\xi \ne 0$), which are the solutions of Eq.~\ref{Eq:Difusao} around $|\mu-\mu_1| \sim 0$.

\begin{figure}[htb]
\begin{center}
\includegraphics[width=.7\columnwidth, angle=-90]{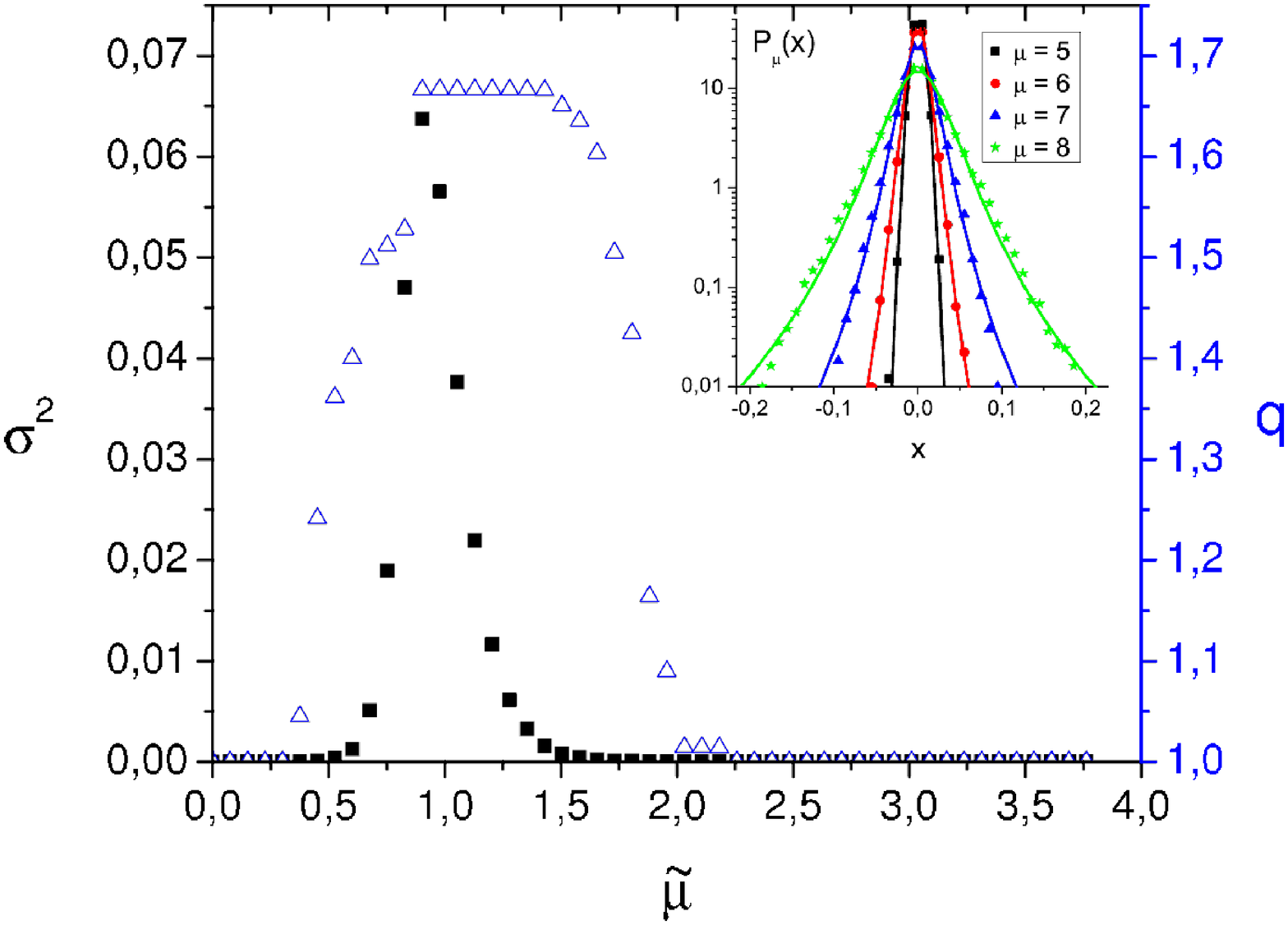}
\caption{On the lefthand side axis one reads the variance $\sigma^2$ ($\bullet$) of curves $P_{\tilde{\mu}}(x)$ (estimator of $\rho$) as function of normalized memory $\tilde{\mu}$.
Observe that $\sigma^2$ diverges near the critical memory.
This region is equivalent to the smallest amplitude, but with maximum dispersion of $P_{\tilde{\mu}}(x)$,  in Fig.~\ref{Fig:SupNormalizada}.
On the righthand side there are the values of $q$ ($\bigtriangleup$) as function of $\tilde{\mu}$.
Notice $q \approx 5/3$ in the region where $\sigma^2$ diverges.
{\bf Inset:} $P_{\mu}(x)$ as a function of the final walker positon for some memory values. 
The displayed curves are $q$-gaussians defined in the text. 
}
\label{Fig:VarQ}
\end{center}
\end{figure}

To fit $P_{\tilde{\mu}}(x)$ to the theoretical model $\rho(x, t)$, we have computed its variance for each $\tilde{\mu}$ and used the maximum likelihood estimation method to find $q$ (Inset of Fig.~\ref{Fig:VarQ}).
One sees that $\sigma^2$ diverges in the transition region presenting a heavy tail if compared to a normal distribution (Fig.~\ref{Fig:SupNormalizada}).
For $\tilde{\mu} \le \log 2$ and $\tilde{\mu} \ge 2.2577$, $P_{\tilde{\mu}}(x)$ shows a gaussian behavior ($q = 1$ and $\xi = 0$).
This is like a random walk model with independent steps and finite variance. 
In one hand, for $\tilde{\mu} < 1$, this is due to the short transient length and cycle period.
The walker excursion is limited to the region near the origin. 
On the other hand, for $\tilde{\mu} > 1$, this occurs since all cycles have a period of $N$.
Thus, after $N$ steps, the walker returns to the starting point and the variance vanishes. 
This implies that the starting point coordinate follows a gaussian pdf.
In the $q$-gaussian model, $\sigma^2$ diverges at $q = 5/3$.
Despite this result being in accordance with the estimated $q$ values, the model does not fit the data for $ 2 \log 2 < \tilde{\mu} < 2.1072$.
In the region $2 \log 2 < \tilde{\mu} < 3 \log 2 $ the distributions are L\'{e}vy $\alpha$-stable, because their tails are heavier than $q$-gaussian ones. 
For $3 \log 2  \le \tilde{\mu} < 2.1072$, the distributions are almost uniform, with a clear central peak.
This behavior is probably due the periodic boundary conditions. 

In short, for $0 \le\tilde{\mu} \le \log 2$ and $\tilde{\mu} > 2.2577$, the tourist follows a traditional diffusion process (gaussian solution). 
For $\log 2 < \tilde{\mu} \le 2 \log 2$,   and $2.1072 \le \tilde{\mu} \le 2.2577$, it is a non-linear superdifusion process (anomalous diffusion with $q$-gaussian solution). 
In this region the anomalous behavior is obtained since the steps are not all independent. 
For $2 \log 2 < \tilde{\mu} < 2.1072$, the process presents another kind of superdiffusion, where the anomalous behavior is strongly due to the fact that the variance is not finite.  
In the region $2 \log 2 < \tilde{\mu} < 3 \log 2$ the distributions are L\'{e}vy $\alpha$-stable.

Finally, the tourist rule can be relaxed to a stochastic walk. 
Thus, the walker goes to nearer cities with greater probabilities.
These probabilities are given by an one-parameter (inverse of the temperature) exponential distribution. 
This situation has been studied for $\mu = 0$~\cite{martinez:1:2004} and $\mu = 1$~\cite{risaugusman:1:2003} and we have detected the existence of a critical temperature separating the localized from the extended regimes. 
It would be interesting to combine both  stochastic movimentation (driven by a temperature parameter) and memory ($\mu$) in the tourist walks to perform full compromise between medium exploration and distance travelled. 


In conclusion, if the walker has critical memory $\mu_1$ the solutions of Eq.~\ref{Eq:Difusao} have infinite variance.
This favors the walker to explore the whole system with minimum displacement, indicating that $\mu_1$ is an optimum memory for an exploratory strategy. 
It is intriguing that a simple deterministic system as this one can present a complex behavior given by the complete fractionary non-linear difusion equation and that a small memory value allows the global medium exploration.


The authors thank N. A. Alves and F. M. Ramos for fruitful discussions.
ASM acknowledges CNPq (305527/2004-5) and FAPESP (2005/02408-0) for support.  


\end{document}